\documentclass[aps,prl,twocolumn,groupedaddress,nofootinbib,showpacs]{revtex4}
\usepackage{graphicx,epsf,color,amsmath}
\usepackage[caption=false]{subfig}
\newif\ifdraft
\drafttrue
\newif\ifpreprint
\preprinttrue

\def\Section#1{\noindent{\it #1---}}

\def\fig#1{Fig.~{\ref{#1}}}

\def\tab#1{Table~{\ref{#1}}}

\def\spa#1.#2{\left\langle#1\,#2\right\rangle}
\def\spb#1.#2{\left[#1\,#2\right]}
\def\tree{{\rm tree}}

\def\E{{\cal E}}

\def\eqn#1{Eq.~(\ref{#1})}

\def\NeqFour{{{\cal N}=4}}

\def\NeqEight{{{\cal N}=8}}

\def\NTwoMC{{{\rm N}^2{\rm MC}}}

\def\NThreeMC{{{\rm N}^3{\rm MC}}}

\def\NSixMCs{{{\rm N}^6{\rm MCs}}}

\def\NkMCs{{{\rm N}^k{\rm MCs}}}

\def\be{\begin{equation}}
\def\ee{\end{equation}}
\def\bea{\begin{eqnarray}}
\def\eea{\end{eqnarray}}
\def\ba{\begin{eqnarray}}
\def\ea{\end{eqnarray}}

\def\tree{{\rm tree}}

\newbox\charbox
\newbox\slabox
\def\s#1{{      
\setbox\charbox=\hbox{$#1$}
\setbox\slabox=\hbox{$/$}
\dimen\charbox=\ht\slabox
\advance\dimen\charbox by -\dp\slabox
\advance\dimen\charbox by -\ht\charbox
\advance\dimen\charbox by \dp\charbox
\divide\dimen\charbox by 2
\raise-\dimen\charbox\hbox to \wd\charbox{\hss/\hss}
\llap{$#1$} }}

\begin{document}

\ifpreprint
\hbox{\hskip0.3cm UCLA/17/TEP/101 \hskip 4.1 cm  NORDITA-2017-3  \hskip4.5 cm
UUITP-01/17}
\fi

\title{Gravity Amplitudes as Generalized Double Copies}

\author{Zvi~Bern${}^{a}$, John Joseph Carrasco${}^b$, Wei-Ming Chen${}^a$, 
Henrik Johansson${}^{c,d}$, Radu Roiban${}^e$ }
\affiliation{
${}^a$Mani L. Bhaumik Institute for Theoretical Physics\\
UCLA Department of Physics and Astronomy\\ 
Los Angeles, CA 90095, USA \\
%
${}^b$Institut de Physique Theorique, CEA-Saclay\\ 
F-91191 Gif-sur-Yvette cedex, France\\
${}^c$Department of Physics and Astronomy, Uppsala University\\
 75108 Uppsala, Sweden \\
${}^d$Nordita, KTH Royal Institute of Technology and Stockholm University\\
Roslagstullsbacken 23, 10691 Stockholm, Sweden \\
${}^e$Institute for Gravitation and the Cosmos, Pennsylvania State University, 
    University Park, PA 16802, USA \\
}

\begin{abstract}
Whenever the integrand of a gauge-theory loop amplitude can be
arranged into a form where the BCJ duality between color and
kinematics is manifest, a corresponding gravity integrand can be
obtained simply via the double-copy procedure.  However, finding such
gauge-theory representations can be challenging, especially at high
loop orders. Here we show that we can instead start from generic
gauge-theory integrands, where the duality is not manifest, and apply
a modified double-copy procedure 
to obtain gravity integrands that include contact
terms generated by violations of dual Jacobi identities.  
We illustrate this with three-, four- and five-loop examples in
$\NeqEight$ supergravity.
\end{abstract}

\pacs{04.65.+e, 11.15.Bt, 11.25.Db, 12.60.Jv \hspace{1cm}}

\maketitle

\Section{Introduction}
Gravity and gauge theories are intimately connected by a double-copy relationship
that was first brought to light by the Kawai--Lewellen--Tye (KLT)~\cite{KLT}
tree-level amplitude relations, and then fleshed out by the Bern--Carrasco--Johansson
(BCJ) duality~\cite{BCJ,BCJLoop} between color and kinematics.
Apart from giving remarkably simple means for obtaining loop-level scattering
amplitudes in a broad class of (super)gravity
theories~\cite{BCJLoop, SimplifyingBCJ,OtherExamples,FundMatter,DoubleCopyTheories,DoubleCopyTheoriesFund},
the duality also addresses the construction of black-hole and other
classical solutions~\cite{ClassicalSolutions} including those
potentially relevant to gravitational-wave
detectors~\cite{RadiationSolutions}, corrections to gravitational
potentials~\cite{Donoghue}, the relation of supergravity symmetries to
gauge-theory ones~\cite{SugraSyms, DoubleCopyTheories, DoubleCopyTheoriesFund}, and the observation of mysterious
``enhanced'' ultraviolet cancellations in certain supergravity
theories~\cite{Enhanced}.  This duality was later found to be
applicable to a wider class of quantum field and string theories~\cite{NonYMBCJ}.
For recent reviews, see Ref.~\cite{Review}.

In the tree-level approximation, manifestly BCJ duality-satisfying
representations of amplitudes are known for any
multiplicity~\cite{BCJTreeProof}.  However, the off-shell duality
remains mysterious despite progress related to the infinite-dimensional Lie algebra underlying BCJ
duality~\cite{UnderlyingKinematicAlgebra} and interesting connections
to gauge symmetries~\cite{BCJGaugeSym}.  At loop level the duality
continues to be a conjecture, with many known
examples~\cite{BCJLoop,SimplifyingBCJ,OtherExamples,FundMatter,DoubleCopyTheories,DoubleCopyTheoriesFund}.
There are also various related double-copy constructions~\cite{NonYMBCJ,CheungAndRemmen}. 

When known,  BCJ-dual representations arguably provide the most efficient
approach to finding loop amplitudes in gauge, gravity and other double-copy
constructible theories. Gauge-theory BCJ representations are usually found
by subjecting an Ansatz to the duality and unitarity constraints.  
However, as the multiplicity and loop order increases,
finding an Ansatz that actually solves the system becomes an increasingly difficult
challenge.
In particular, no BCJ form of the five-loop four-point amplitude of
$\NeqFour$ super-Yang--Mills (sYM) theory has yet been found, although
a BCJ-dual $1/2$-BPS five-loop form factor in this theory has recently
been found~\cite{FiveLoopFormFactor}.
In this Letter, we explain how to apply the duality to construct gravity
integrands from generic (not manifestly BCJ-dual) 
gauge-theory integrands.

\begin{figure}[t]
\includegraphics[clip,scale=0.3]{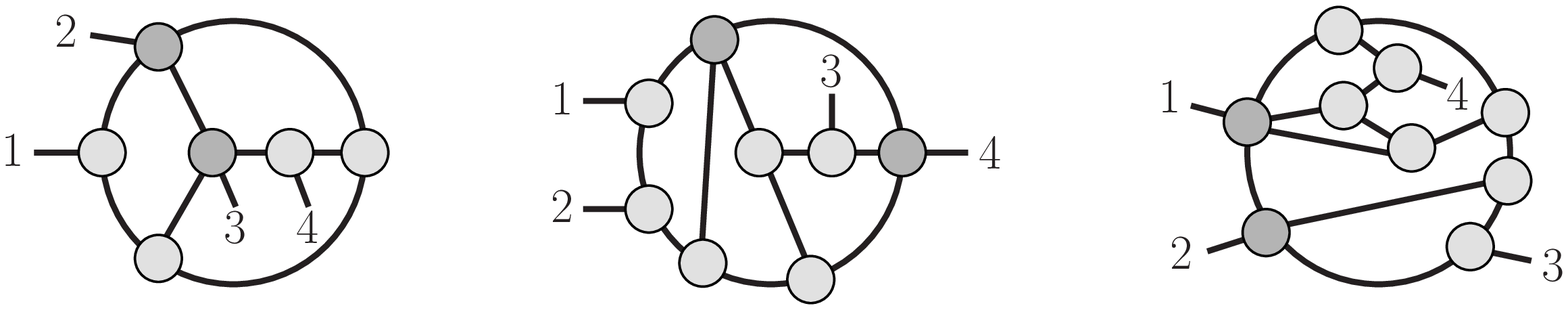}
\vskip -.2 cm 
\caption[a]{Sample N$^k$-maximal cuts at three, four and five loops.
  Exposed lines are all on shell.}
\label{CutSamplesFigure}
\end{figure}

\begin{figure}[t]
\centerline{\epsfxsize 3.4 truein \epsfbox{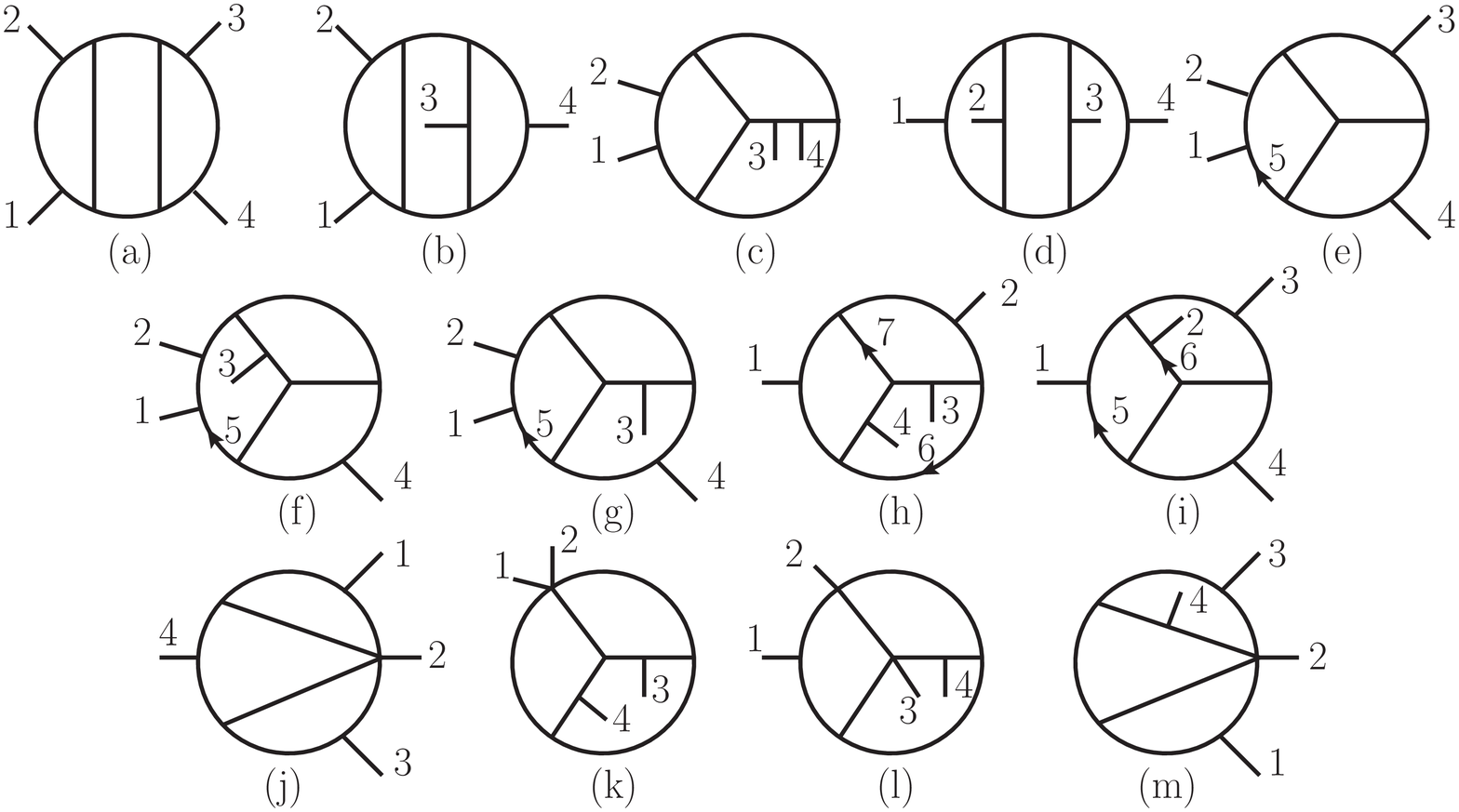}}
\vskip -.2 cm 
\caption[a]{\small Diagrams (a)-(i) define the three-loop four-point amplitude of $\NeqFour$ sYM theory;
diagrams (j)-(m) are the additional contact terms needed for $\NeqEight$ supergravity.}
\label{ThreeLoopDiagramsFigure}
\end{figure}

\Section{Contact Terms from BCJ Duality}
Our derivation of gravity amplitudes uses the method of maximal
cuts~\cite{MaximalCutMethod}, a refinement of generalized
unitarity~\cite{GeneralizedUnitarity}.  In this method, amplitudes are
constructed from cuts that reduce integrands to sums of products of
tree-level amplitudes, as illustrated in \fig{CutSamplesFigure}.  The
cuts are organized according to the number $k$ of propagators that
remain off shell.  We first find an expression whose maximal ($k=0$)
cuts (MCs) are correct, then correct it such that all next-to-maximal
($k=1$) cuts (NMCs) are correct and systematically proceed through the
N$^k$MCs, until no further corrections exist.  The maximal $k$ depends
on the power counting of the theory and on choices made at earlier
levels. The corrections coming from N$^k$MCs are assigned to
contact terms corresponding to each cut.  For example, the first cut
in \fig{CutSamplesFigure} determines the double-contact diagram (l) in
\fig{ThreeLoopDiagramsFigure}.  The contact terms are taken off shell
in a manner that preserves the diagram symmetry.  This process
introduces an ambiguity that can then be absorbed into changes in
subsequent $(k+1)$-level contact terms.

Generically, an $L$-loop $m$-point gravity amplitude can be organized in
terms of diagrams that have cubic and higher-point (contact term) vertices,
\begin{equation}
{\cal M}^{(L)}_m =  i^{L} \Bigl(\frac{\kappa}{2}\Bigr)^{m-2 +2L}
\sum_{j\in \Gamma_L} \!\!
\int \frac{d^{LD} p}{ (2 \pi)^{LD}}
\frac{1}{S_j}  \frac {N_j}{\prod_{\lambda_j} \! {d^{(\lambda_j)}_j}},
\label{LoopGravity}
\end{equation}
where $\Gamma_L$ is the set of all $L$-loop $m$-point graphs with
labeled external lines, $\lambda_j$ labels the edges of graph $j$,
and $1/d^{(\lambda_j)}_j$ are the corresponding propagators.  For
example, for the three-loop four-point $\NeqEight$ amplitude, the diagrams needed
are given in \fig{ThreeLoopDiagramsFigure}.  The symmetry factors
$S_j$ remove overcounts arising from automorphisms of each graph.
The integration is over $L$ independent $D$-dimensional loop momenta.
The gravity numerators $N_j$ will be obtained through our double-copy
procedure, in a manner that respects the graph's symmetries.

BCJ duality~\cite{BCJ,BCJLoop} is manifest if the kinematic
numerators $n$ of a gauge-theory amplitude's graphs have the same algebraic
properties as its color factors~$c$. The basic Jacobi identity of the
gauge-group structure constants can be embedded in arbitrary multiloop
diagrams and leads to relations between the color factors $c$ of
triplets $\{A,B,C\}$ of graphs.  
For generic theories with only fields in the adjoint representation, the duality implies the 
functional relations 
\begin{equation}
c_A + c_B + c_C = 0 \;  \longleftrightarrow \;  n^\text{BCJ}_A + n^\text{BCJ}_B + n^\text{BCJ}_C = 0 \,,
\label{BCJDuality}
\end{equation}
for all such triplets.
Generalized gauge transformations---shifts of $n$ that cancel in the
amplitude because of color Jacobi relations--- bring the kinematic
numerators to this form.  Numerators that satisfy
\eqn{BCJDuality} lead directly to gravity integrands by replacing
color factors of the gauge-theory amplitude with the kinematic
numerators of a second gauge-theory amplitude~\cite{BCJ,BCJLoop,Square}.

At loop level, the functional relations \eqref{BCJDuality} 
are typically solved through Ans\"atze obeying additional simplifying
assumptions~\cite{SimplifyingBCJ}. While the resulting expressions are
compact, sufficiently unconstrained Ans\"atze  can be prohibitively large. 
This motivates us to find an alternative efficient double-copy construction of gravity 
integrands that evades the need for explicit
duality-satisfying representations of gauge-theory amplitudes.

The starting point of our construction is a ``naive double copy'' of 
amplitudes of two possibly distinct gauge theories, written in terms of 
cubic diagrams with numerator factors $n_i$ and $\tilde n_i$ for which 
duality (\ref{BCJDuality}) is not manifest:
\be
N_i = n_i \tilde n_i  ~~\text{(cubic only)}\, .
\label{naive2copy}
\ee
This naive double copy automatically
satisfies all the gravity MCs and NMCs, because BCJ duality always
holds for on-shell four-point tree amplitudes~\cite{BCJ}.  However,
because BCJ duality is not manifest, Eq.~\eqref{naive2copy} is not the
complete answer, as can be checked by evaluating N$^2$MCs. We need a
systematic determination of the contact-term corrections that lead to
the correct $\NkMCs$, matching the result that could be independently
obtained by less efficient methods, such as applying KLT relations
directly on the cuts.

The key observation is that the additional contact contributions should
be related to the violation of the kinematic Jacobi relations~\eqref{BCJDuality}  by
the gauge-theory amplitude numerators. Together with generalized gauge
invariance and properties of BCJ numerators in the generalized cuts,
this observation provides the building blocks for the construction of
the missing terms.  To describe their construction we need a labeling
of a general cut ${\cal C}$, made of $q$ factors of $4 \le m$-point tree
amplitudes. We choose an ordering, $1,\dots,q$, of these amplitude
factors, an ordering of the graphs contributing to each such factor
and label numerators by the labels of the graph in each amplitude
factor, $n_{i_1, i_2, ..., i_q}$ with $i_1$ running over the graphs in
the first amplitude factor, {\it etc.}  For example, the numerators of
the nine diagrams (some of which can vanish) appearing in the first
cut in \fig{CutSamplesFigure} are labeled by $n_{i_1, i_2}$, where
$i_1$ and $i_2$ each runs over the three graphs in a four-point tree
amplitude.

For every propagator of every graph contributing to a generalized cut there is a
kinematic Jacobi relation. Choosing
an ordering for the propagators in every graph, we define the
violation of the kinematic Jacobi relation on the $\lambda_A$-th propagator of
graph $A$ of $v$-th amplitude factor:
\begin{eqnarray}
&& \hskip -.3 cm 
J_{i_1,\dots, i_{v-1}, \{A, \lambda_A\} , i_{v+1},\dots, i_q}  =
n_{i_1,\dots, i_{v-1}, A, i_{v+1},\dots, i_q} \cr
&&\quad\null
+n_{i_1,\dots, i_{v-1}, B, i_{v+1},\dots, i_q} 
+n_{i_1,\dots, i_{v-1}, C, i_{v+1},\dots, i_q} \,, \hskip .5 cm
\label{Jdef}
\end{eqnarray}
where graphs $B$ and $C$ are connected to graph $A$ by the color
Jacobi relation on the $\lambda_A$-th propagator of graph $A$.  (The relative
signs between terms should match those of the corresponding color
Jacobi relation.)  We can define, in the obvious way, violations of
multiple Jacobi relations; they are linear combinations of these.

Not all such $J$s are independent. First, there is a triple over-count,
since the same $J$ can be defined for each of the three diagrams
connected by a Jacobi relation. 
Furthermore, there are linear relations, some from the definition (\ref{Jdef}) 
of $J$ in terms of $n$ and some due to BCJ amplitude
relations~\cite{BCJ}.  Tree-level examples of these latter $J$
relationships are derived in
Refs.~\cite{HenryConstraints,PierreConstraints}.

With this notation, the generalized gauge shift, $\Delta$, that relates arbitrary 
kinematic numerators $n$ to BCJ ones~is
\begin{eqnarray}
n_{i_1, i_2, ...i_q} &=& n_{i_1, i_2, ...i_q}^\text{BCJ} 
  + \Delta_{i_1, i_2, ...i_q}\,,
\cr
\Delta_{i_1, i_2, ...i_q} &=& \sum_{v} \sum_{j} d_{i_v}^{(v,j)} 
\alpha^{(v,j)}_{i_1,\dots {\hat \imath_v} ,\dots i_q}\,,
\label{gauge_transf}
\end{eqnarray}
where the hat notation indicates that the index is omitted,
and $j$ runs over the labels of 
the ordered set of inverse propagators of the graph
$i_v$ of the $v$-th amplitude factor and $d_{i_v}^{(v,j)}$ is the
$j$-th element of this set.  The $\Delta$ are constrained 
so they do not alter the gauge-theory cut integrand~\cite{BCJ,BCJLoop,Square}
\be
\sum_{i_1,\dots,i_q} \frac{\Delta_{i_1, i_2, ...i_q} c_{i_1, i_2, ...i_q} }
{D_{i_1}\dots D_{i_q}} = 0 \ ,
\label{gt}
\ee 
where $D_{i_v}$ is the product of all inverse propagators of the
graph $i_v$ in the $v$-th amplitude factor. 

Using Eq.~\eqref{gauge_transf}, the gravity cut is given by
\begin{eqnarray}
{\cal C}_\text{G} &=& \sum_{i_1,\dots,i_q} \frac{(n_{i_1, i_2, ...i_q} ^\text{BCJ})^2}
  {D_{i_1}\dots D_{i_q}}
  =
  \sum_{i_1,\dots,i_q} \frac{n_{i_1, i_2, ...i_q} ^2}
  {D_{i_1}\dots D_{i_q}} +{\cal E}_\text{G}\,,
\cr
{\cal E}_\text{G}&=& - \sum_{i_1,\dots,i_q} \frac{\Delta_{i_1, i_2, ...i_q}^2}
  {D_{i_1}\dots D_{i_q}} \,.
\label{cut_exp}
\end{eqnarray}
For simplicity we have taken the two gauge-theory numerators to be identical.
The key to the cancellation of the $n^\text{BCJ}\cdot \Delta $ cross 
terms is Eq.~\eqref{gt}, given that the $n^\text{BCJ}$ satisfy Eq.~\eqref{BCJDuality}.  We
stress that this argument relies only on the existence, but not
explicit construction, of tree-level BCJ representations used in the
generalized cuts.  

To express ${\cal E}_\text{G}$ in terms of $J$s requires inverting, on
a case by case basis, the relations $J(\Delta)$ obtained by plugging
Eqs.~\eqref{gauge_transf} into Eq.~\eqref{Jdef}.  Since not all $J$s are
independent, only some gauge shifts can be determined. The remaining
ones preserve the BCJ form of the gauge-theory cut. The resulting
expression in a complete $J$-basis superficially has spurious
singularities; they may be eliminated explicitly by using the
remaining gauge freedom.

We illustrate the general construction described here by discussing
in some detail the N$^2$MCs made of two four-point
amplitudes. The numerators are labeled as $n_{i_1, i_2}$ where $i_1$
and $i_2$ run over the three graphs in the first and second four-point
amplitude, respectively. Each graph has a single propagator; 
the second upper index on inverse propagators is therefore 
redundant so we do not include it.
The generalized gauge transformation \eqref{gauge_transf} is 
\be
\Delta_{i_1,i_2} = d^{(1)}_{i_1} \alpha^{(1)}_{i_2} + d^{(2)}_{i_2} \alpha^{(2)}_{i_1}\,.
\label{4x4Delta}
\ee
After use of momentum conservation $\sum_{i_1} d_{i_1}^{(v)}=0$,
this gives the violations of kinematic Jacobi relations as
\bea
J_{\{u_1, 1\}, i_2} &\equiv& \sum_{i_1} n_{i_1, i_2}
  = d^{(2)}_{i_2}  \sum_{i_1} \alpha^{(2)}_{i_1}\,,
\nonumber \\
J_{i_1, \{u_2, 1\}} &\equiv& \sum_{i_2} n_{i_1, i_2} 
 = d^{(1)}_{i_1}  \sum_{i_2} \alpha^{(1)}_{i_2}\, .
\label{4x4J}
\eea
The threefold degeneracy of $J$ implies independence on the labels $u_1$ or
$u_2$.  We see that only particular combinations of gauge shifts $\alpha^{(v)}_i$ are determined.
We also note that $J_{i_1, \{u_2, 1\}}/d^{(1)}_{i_1} $ and
$J_{\{u_1, 1\}, i_2} /d^{(2)}_{i_2}$ are independent of the graph in
the second and first amplitude, respectively.

Combining Eqs.~\eqref{cut_exp}, \eqref{4x4Delta} and \eqref{4x4J}, the additional contact term 
in a N$^2$MC with two four-point amplitudes is
\be
\E^{4\times 4}_\text{G} =
- 2 \sum_{i_1,i_2} \alpha^{(1)}_{i_1} \alpha^{(2)}_{i_2}  =
- 2 \frac{J_{\{1, 1\}, 1}  J_{1, \{1, 1\}} }{d^{(1)}_{1} d^{(2)}_{1}} \, .
\label{Contact4x4}
\ee
The two denominators cancel against the numerator,
yielding a local expression.

Similar but more involved analysis gives formulae correcting
any cut of a naive double copy.  Unlike the example above, the
nontrivial constraints between $J$s, as well as the requirement that
generalized gauge transformations should not modify gravity
amplitudes, require nontrivial disentangling.  At the $\NTwoMC$
level the extra term corresponding to a five-point contact term is
\be
{\cal E}^{5}_\text{G} = 
-  \frac{1}{3}\sum_{i=1}^{15} \frac{J_{\{i, 1\}} {J}_{\{i, 2\}}}{d^{(1,1)}_{i} d^{(1,2)}_{i}} \,,
\label{Contact5}
\ee
where the sum runs over all 15 cubic graphs of the five-point
amplitude and $d_{i}^{(1,1)}$  and $d_{i}^{(1,2)}$ are the two propagators of graph $i$.
This formula is a symmetric loop-generalization of the one given in
Ref.~\cite{PierreConstraints} for tree amplitudes.

At the $\NThreeMC$ level, the extra terms are more involved. For example, the 
correction terms relevant to three four-point amplitude cuts are: 
\bea
 \label{Contact4x4x4}
&& \hskip -.7 cm 
{\cal E}^{4\times4\times4}_\text{G} = 2\, \frac{J^{(1)}_{1,1} J^{(2,3)}_1 + J^{(2)}_{1,1} J^{(1,3)}_1 
         + J^{(3)}_{1,1} J^{(1,2)}_1 } {d^{(1)}_1 d^{(2)}_1 d^{(3)}_1}
\\
&-&\!\! \sum_{i_3} \frac{2J^{(1)}_{1, i_3} J^{(2)}_{1, i_3} } 
                  {d^{(1)}_1 d^{(2)}_1 d^{(3)}_{i_3}}
- \sum_{i_2} \frac{2J^{(1)}_{i_2, 1} J^{(3)}_{1, i_2} }
                  {d^{(1)}_1 d^{(2)}_{i_2} d^{(3)}_{1}} 
-  \sum_{i_1} \frac{2J^{(2)}_{i_1, 1} J^{(3)}_{i_1, 1} } 
                  {d^{(1)}_{i_1} d^{(2)}_1 d^{(3)}_{1}} \,,
\nonumber                  
\eea
where we used the shorthand notations $J^{(1)}_{1, i_3}\equiv J_{\{1, 1\}, 1, i_3}$, {\it etc.}, and
$J^{(1,2)}_{i_3} \equiv \sum_{i_1} J_{i_1,\{1, 1\}, i_3}$, {\it etc.}
As in \eqn{4x4Delta}, we have suppressed the second upper index on $d_i^{(v,j)}$ 
because it takes a single value.
We have also derived general formulae for cuts with $4\times 5$ and
$6$-point amplitude factors, which we will present together with the
$\NeqEight$ supergravity five-loop four-point
integrand~\cite{NextPaper}.

One subtlety is that, in special cuts, momentum conservation can
force on shell an internal propagator of a tree amplitude, leading to a 1/0 divergence.  
This is associated with bubble on external leg or
tadpole diagrams, which in dimensionally regulated massless theories
integrate to zero.  In $\NeqEight$ supergravity the simplest prescription
is to take such contributions to vanish whenever a corresponding
numerator vanishes and, if the cancellation occurs between terms
(which can leave finite pieces behind) to take advantage of the
asymmetry of the formul\ae{} to choose a labeling that avoids this
situation.

Locality and dimension counting imply that contact terms become
simpler as the level increases.  For example,  the contact-term numerators 
at the $\NSixMCs$ level in the five-loop four-point amplitude of $\NeqEight$ 
supergravity are just a linear combination of $s^2$, $st$ and $t^2$. 
Thus, in practical calculations, it is more
efficient to determine the high-level contact terms by numerically
evaluating the generalized cuts.

\begin{figure}[t]
\centerline{\epsfxsize 2.5 truein \epsfbox{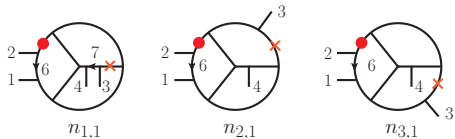}}
\vskip -.2 cm 
\caption[a]{\small The three diagrams whose kinematic numerators
  contribute to $J_{\{1,1\},1}$.  The thick shaded (red) cross marks
  the off-shell legs participating in the dual Jacobi relation.
  The shaded (red) dot indicates the off-shell leg of the second
  amplitude factor. }
\label{ThreeLoopJFigure}
\end{figure}

\begin{table}
\caption{A non-BCJ form of the three-loop four-point $\NeqFour$ sYM
  diagram numerators from Ref.~\cite{ManifestPower3}.  
 We define $\tau_{ij} = 2 p_i \cdot p_j$, $s= (p_1
 + p_2)^2$, $t = (p_2+p_3)^2$ and $u = (p_1+p_3)^2$. 
}
\label{NonBCJNumeratorTable} 
\vskip .2 cm 
\begin{tabular}{||c|c||}
\hline
Graph & $\NeqFour$ sYM numerators.  \\
\hline
(a)-(d)& $s^2$\\
\hline
(e)-(g)& $s (p_5^2 + \tau_{45})$\\
\hline
(h)&$ s(\tau_{26} + \tau_{36}) -
      t (\tau_{17}+ \tau_{27}) + s t $ \\
\hline
(i)& $  s (p_5^2 + \tau_{45}) - t (p_5^2 + \tau_{56} + p_6^2) -
        (s - t) p_6^2/3 $\\
\hline
\end{tabular}
\end{table}

\Section{Examples}
The three-loop four-point amplitude of $\NeqEight$ supergravity is
well studied~\cite{ThreeloopN8,ManifestPower3,BCJLoop,SimplifyingBCJ}
and~serves as a useful illustration. We will reconstruct it here from
the corresponding $\NeqFour$ sYM amplitude of
Ref.~\cite{ManifestPower3} whose numerators are displayed in
\tab{NonBCJNumeratorTable} with the momentum labeling in
\fig{ThreeLoopDiagramsFigure}(a)-(i) (corresponding to the one of
Ref.~\cite{BCJLoop}).  
As in the normalization of Ref.~\cite{BCJLoop}, an overall factor of $s t
A_4^\tree$ is removed.  Following our
procedure, the $\NeqEight$ supergravity numerators of diagrams
(a)--(l) are squares of the corresponding $\NeqFour$ sYM ones:
\begin{equation}
N^{\NeqEight}_{(x)} =n_{(x)}^2  \,, \hskip 1 cm  x \in \{\rm a,\dots,  i\} \,.
\end{equation}

Contact diagrams can appear only at the $\NTwoMC$ level.  
There are a total of 62 possible independent such contact terms. 
Of these, all but the four diagrams
(j)-(m) in \fig{ThreeLoopDiagramsFigure} vanish.  As an example,
consider the contact diagram in \fig{ThreeLoopDiagramsFigure}(l),
composed of two four-point vertices.  We obtain it from
\eqn{Contact4x4}.
First, we identify the nine cubic diagrams that contribute to it (some
are vanishing) and pick one whose numerator we label as $n_{1,1}$; we
choose diagram (c) in \fig{ThreeLoopDiagramsFigure}. The two 
$J$-functions are calculated by relabeling the appropriate
numerators to the labels of \fig{ThreeLoopJFigure}.  For example, 
$J_{\{u_1, 1\},1}$ is obtained from the $\NeqFour$ sYM
numerators of the three diagrams shown in \fig{ThreeLoopJFigure}, 
\be
n_{1,1} = s^2, \hskip .15 cm n_{2,1} = s (t+ \tau_{26} +\tau_{36}), 
\hskip .15 cm
 n_{3,1} = s (u - \tau_{36}) \,,
\ee 
corresponding to relabeling of diagrams (c) and (g) in
\fig{ThreeLoopDiagramsFigure}.  Summing and applying momentum
conservation gives $J_{\{1, 1\}, 1} = s \tau_{26}$. Similarly, $J_{1,
  \{1, 1\}} = s \tau_{37}$.  With these labels, the two off-shell
inverse propagators are $\tau_{26}$ and $\tau_{37}$, so that from
\eqn{Contact4x4} the $\NeqEight$ supergravity contact term numerator
for diagram (l) is
\begin{equation}
 N^{\NeqEight}_{\rm (l)} = -2\frac{J_{\{1, 1\}, 1} J_{1, \{1, 1\}}} 
 {\tau_{26} \tau_{37}} = -2 s^2\,.
\end{equation}
The other three independent contact terms corresponding to diagrams
(j), (k) and (m), can similarly be obtained from \eqn{Contact5},
with the result 
\be
N^{\NeqEight}_{\rm (j)} = - {\textstyle \frac{1}{9}} (s-t)^2\,, \hskip .4 cm 
N^{\NeqEight}_{\rm (k)} = N^{\NeqEight}_{\rm (m)} = -2s^2\,.
\end{equation}
All nonvanishing contact terms are relabelings of these. 

We have also computed the four-loop four-point amplitude of $\NeqEight$
supergravity using the contact-term method described above.  The
results are included as a mathematica
attachment~\cite{AttachedFile}.  Power counting dictates that this
$\NeqEight$ amplitude can have no contact terms beyond level $k=4$, which we 
checked explicitly.

Generating contact-term diagrams by collapsing the propagators of the
cubic-contributions in all possible ways, we find the result is surprisingly
simple. The vast majority of contacts, 2353 of 2621, vanish outright
due to vanishing $J$'s. (In this count, we drop cuts where a leg of a
tree-amplitude is directly sewn to another one of the same tree, since
these do not appear in $\NeqEight$ supergravity.)  Even the
nonvanishing 268 contact terms are remarkably simple.  For example, as
for the three-loop cut (l), we evaluated the four-loop contact term
given by the second cut in \fig{CutSamplesFigure} (corresponding to
the 48th $\NTwoMC$ in the attachment \cite{AttachedFile}) using
\eqn{Contact4x4} and found that its numerator is $(-2s^4)$.  All
remaining contact terms are included in the attached
file~\cite{AttachedFile}.

By five loops, even promising methods may prove ineffective due to a
combinatorial proliferation of terms.  We therefore perform extensive
checks at five loops to ensure that the methods presented here remain
practical.  For example, starting from the $\NeqFour$ sYM expression
in Ref.~\cite{FiveLoopN4}, we find that the overwhelming fraction of
contact terms through $\NSixMCs$ are zero, such as the rather
nontrivial third cut in \fig{CutSamplesFigure}.  This is consistent
with lower loops and enormously simplifies the construction and
structure of the $\NeqEight$ supergravity five-loop four-point
amplitude, to be described elsewhere~\cite{NextPaper}.

\Section{Conclusions and Outlook}
Some open problems remain.
The power counting of the gravity integrands given by our modified  
double-copy construction depends on the choices of numerators in the sYM amplitude; generic 
representations of the latter typically lead to higher-than-optimal power 
counting of the former.
For example, the known five-loop $\NeqFour$ sYM integrand is of this type.  
Gauge-theory integrands designed to minimize the power count in the 
double copy, in particular of its naive part, are therefore desirable; 
finding them is an important problem.

Although we focused here on $\NeqEight$ supergravity and $\NeqFour$
sYM, the construction generalizes in the obvious way to different
gauge and other theories with adjoint matter which 
obey BCJ duality, and thus to all gravitational and non-gravitational~\cite{NonYMBCJ} double-copy
theories obtained from them. 
If the two theories in the double copy are distinct, the contact terms
are obtained by simply replacing in our expressions $J_i J_j
\rightarrow (J_i \widetilde{J}_j+\widetilde{J}_i J_j)/2$, where $J$
and ${\widetilde J}$ are the violations of the kinematic Jacobi
relations in each theory.

Similar ideas to the ones presented in
this Letter should hold in all double-copy theories whose single-copies include
fields in the fundamental representation of the gauge group~\cite{FundMatter, DoubleCopyTheoriesFund}. 
Our results suggest that it may be possible to generically convert any gauge-theory 
classical solution to a gravitational one without choosing special generalized gauges.
We expect that the ideas presented in this paper will be useful not
only for investigating the ultraviolet behavior of perturbative
quantum gravity but also for understanding general physical properties
of gravity theories.

\vskip .15 cm 

\Section{Acknowledgments}
We thank Jacob Bourjaily, Alex Edison, David Kosower, Enrico Hermann and
Jaroslav Trnka for many useful and interesting discussions.  This work
is supported by the Department of Energy under Award Numbers
DE-SC0009937 and DE-SC0013699. J.~J.~M.~C. is supported by the
European Research Council under ERC-STG-639729, {\it preQFT: Strategic
  Predictions for Quantum Field Theories}.  The research of H.~J. is
supported in part by the Swedish Research Council under grant
621-2014-5722, the Knut and Alice Wallenberg Foundation under grant
KAW 2013.0235, and the Ragnar S\"{o}derberg Foundation under grant
S1/16.  W.-M.~C. thanks Mani L. Bhaumik for his generous support.


\end{document}